\documentclass[twocolumn, prl, a4paper, 10pt, superscriptaddress, floatfix]{revtex4-1}
\usepackage{siunitx}
\usepackage{graphicx}
\usepackage{xcolor}
\usepackage{physics}
\usepackage{nicefrac}
\usepackage{bm}
\usepackage{float}
\usepackage[utf8]{inputenc}
\usepackage[colorlinks, citecolor=blue, linkcolor=black]{hyperref}
\usepackage{ulem}
\usepackage{booktabs}
\usepackage{multirow}
\usepackage{xspace}

\newcommand{\VB}{V$_\text{B}^-$\xspace}

\begin{document}

\title{Isotopic control of the boron-vacancy spin defect in hexagonal boron nitride}

\author{T. Clua-Provost}
\thanks{Contributed equally to this work.}
\author{A. Durand}
\thanks{Contributed equally to this work.}
\author{Z. Mu}
\affiliation{Laboratoire Charles Coulomb, Universit\'e de Montpellier and CNRS, 34095 Montpellier, France}
\author{T. Rastoin}
\affiliation{Laboratoire Charles Coulomb, Universit\'e de Montpellier and CNRS, 34095 Montpellier, France}
\author{J. Frauni\'e}
\affiliation{Universit\'e de Toulouse, INSA-CNRS-UPS, LPCNO, 135 Avenue Rangueil, 31077 Toulouse, France}
\author{E. Janzen}
\author{H. Schutte}
\author{J.~H.~Edgar}
\affiliation{Tim Taylor Department of Chemical Engineering, Kansas State University, Kansas 66506, USA}
\author{G. Seine}
\author{A. Claverie}
\affiliation{CEMES-CNRS and Universit\'e de Toulouse, 29 rue J. Marvig, 31055 Toulouse, France}
\author{X.~Marie}
\author{C.~Robert}
\affiliation{Universit\'e de Toulouse, INSA-CNRS-UPS, LPCNO, 135 Avenue Rangueil, 31077 Toulouse, France}
\author{B. Gil}
\author{G. Cassabois}
\author{V. Jacques}
\email{vincent.jacques@umontpellier.fr}
\affiliation{Laboratoire Charles Coulomb, Universit\'e de Montpellier and CNRS, 34095 Montpellier, France}

\begin{abstract}
We report on electron spin resonance (ESR) spectroscopy of boron-vacancy (\VB) centers hosted in isotopically-engineered hexagonal boron nitride (hBN) crystals. We first show that isotopic purification of hBN with $^{15}$N yields a simplified and well-resolved hyperfine structure of \VB centers, while purification with $^{10}$B leads to narrower ESR linewidths. These results establish isotopically-purified h$^{10}$B$^{15}$N crystals as the optimal host material for future use of \VB spin defects in quantum technologies. Capitalizing on these findings, we then demonstrate optically-induced polarization of $^{15}$N nuclei in h$^{10}$B$^{15}$N, whose mechanism relies on electron-nuclear spin mixing in the \VB ground state. This work opens up new prospects for future developments of spin-based quantum sensors and simulators on a two-dimensional material platform.
\end{abstract} 
\date{\today}

\maketitle

The negatively-charged boron-vacancy (\VB) center in hexagonal boron nitride (hBN) is a promising quantum system for developing spin-based quantum technologies on a two-dimensional (2D) material platform~\cite{doi:10.1080/23746149.2023.2206049}. The electronic spin state of this point defect can be initialized and readout by optical means under ambient condition~\cite{Gottscholl2020}, even in the limit of atomically-thin hBN layers~\cite{DurandArxiv}, offering interesting prospects for quantum sensing and imaging with unprecedented proximity to the sample being probed~\cite{GottschollNatCom2021,GaoStrain2022,IgorStrain2022,Tetienne2023,Du2021,PhysRevApplied.18.L061002,rizzato2022extending}. In addition, the ability to initialize and control strongly interacting \VB spin defects~\cite{Gong2023}, as well as their neighboring nuclear spins in the hBN lattice~\cite{Gao2022}, opens up appealing perspectives for quantum simulations by providing new ways of exploring many-body quantum phenomena in an intrinsically 2D solid-state platform~\cite{Cai2013,Davis2023}.\\
\indent In hBN, each site of the honeycomb lattice is occupied by an atom with a non-zero nuclear spin, forming a highly dense nuclear spin environment. Boron has two stable isotopes, $^{11}$B ($I=3/2$, $80\%$ natural abundance~(n.a.)) and $^{10}$B ($I=3$, $20\%$ n.a.), while nitrogen occurs either as $^{14}$N ($I=1$, $99.6\%$ n.a.) or $^{15}$N ($I=1/2$, $0.4\%$~n.a.). In hBN crystals with a natural isotopic content, the electron spin resonance (ESR) of \VB centers is characterized by a complex seven-line structure resulting from the hyperfine interaction with the three $^{14}$N nuclei placed in the first neighboring lattice sites~\cite{Gottscholl2020}. The hyperfine coupling with the second-neighbors boron atoms rather leads to an overall broadening of each ESR line~\cite{haykal2021decoherence}.\\
\indent In this work, we investigate \VB centers in isotopically-engineered hBN crystals. We first show that isotopic purification with $^{15}$N yields a simplified and well-resolved hyperfine structure of \VB centers, while purification with $^{10}$B leads to narrower ESR linewidths. These results establish isotopically-purified h$^{10}$B$^{15}$N crystals as the optimal host material for future use of \VB spin defects in quantum technologies. We then demonstrate dynamic polarization of $^{15}$N nuclei via optical pumping of \VB centers  in h$^{10}$B$^{15}$N. Our analysis indicates that nuclear polarization is mediated by electron-nuclear spin mixing in the ground state of the \VB center, with a maximal efficiency of $\sim 30\%$, which is mainly limited by the unequal diagonal components of the hyperfine tensor.\\
\indent We rely on millimeter-sized hBN crystals with controlled isotopic content, which are synthesized through the metal flux growth method described in Ref.~\cite{EdgarArxiv}. The source materials are (i) boron powders isotopically enriched with either $^{11}$B ($99.4\%$) or $^{10}$B ($99.2\%$), and (ii) a nitrogen gas featuring either a natural $^{14}$N content or isotopic enrichment with $^{15}$N ($>99\%$). Raman and deep-UV spectroscopy recently confirmed that such a growth method provides high-quality hBN crystals with different configurations of boron and nitrogen isotopes, namely h$^{10}$B$^{14}$N, h$^{11}$B$^{14}$N, h$^{10}$B$^{15}$N, and h$^{11}$B$^{15}$N~\cite{EdgarArxiv}. Few tens of nanometers thick hBN flakes were mechanically exfoliated from these isotopically-purified crystals and transferred on a SiO$_2$/Si substrate. The exfoliated flakes were then implanted with nitrogen at $30$~keV energy with a dose of $10^{14}$~ions/cm$^2$ to create \VB centers. The implantation energy is such that nitrogen atoms stop in the SiO$_2$/Si substrate after scattering through the entire thickness of the hBN flakes, thus producing vacancies without modifying their isotopic content.\\
 \begin{figure*}[t!]
  \centering
  \includegraphics[width = 17.2cm]{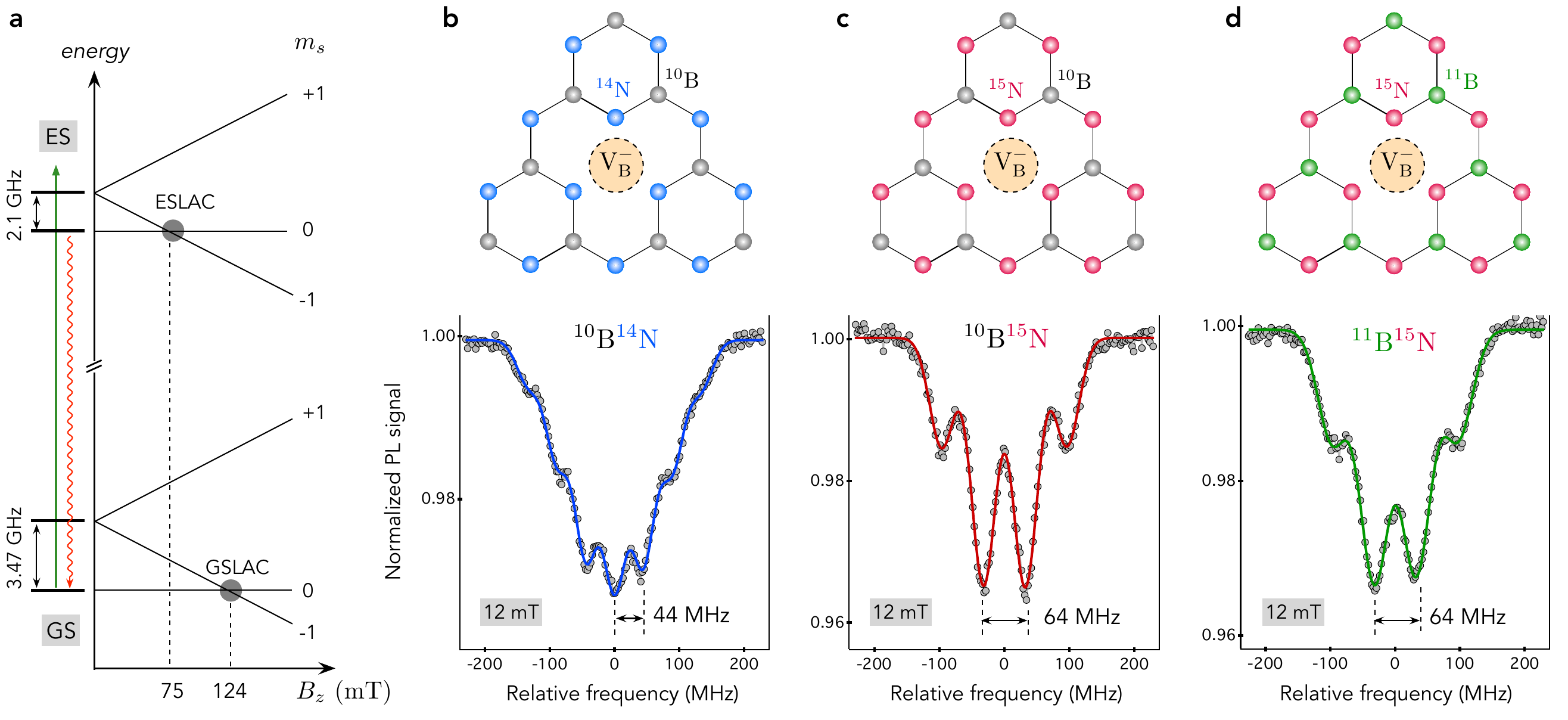}
  \caption{(a) Simplified energy level structure of the \VB center in hBN showing the electron spin sublevels $m_s=0,\pm1$ in the ground (GS) and excited states (ES), as well as their evolution with a static magnetic field $B_z$ applied along the c-axis of hBN. Anticrossing of $m_s=0$ and $m_s=-1$ spin sublevels occurs for $B_z\sim 75$~mT in the excited state (ESLAC) and for $B_z\sim 124$~mT in the ground state (GSLAC). (b-d) Hyperfine structure of the \VB center in (b) h$^{10}$B$^{14}$N, (c) h$^{10}$B$^{15}$N, and (d) h$^{11}$B$^{15}$N crystals. Experiments are done at $B_z=12$~mT with identical microwave and optical excitation powers. The solid lines are data fitting with a sum of Gaussian functions from which we extract the hyperfine splitting $\mathcal{A}_{zz}$ and the full width at half maximum (FWHM) $\Delta$ of the ESR lines.}
  \label{fig1}
\end{figure*}
\indent A simplified energy level structure of the \VB center is sketched in Fig.~\ref{fig1}(a). The ground state is a spin triplet ($S=1$), with a zero-field splitting $D_{\rm g}\sim3.47$~GHz between the spin sublevels $m_s = 0$ and $m_s = \pm1$, where $m_s$ denotes the electron spin projection along the c-axis of the hBN crystal~\cite{Gottscholl2020}. Under optical illumination, the \VB center can be promoted to an excited state, which is also a spin triplet with a zero-field splitting parameter $D_{\rm e}\sim2.1$~GHz~\cite{mathur2021excitedstate,yu2021excitedstate,baber2021excited,mu2021excitedstate}. Relaxation to the ground state occurs either through the emission of a broadband photoluminescence (PL) signal in the near infrared, or via non-radiative channels involving transitions to metastable singlet states [not shown in Fig.~\ref{fig1}(a)]. While optical transitions are mainly spin conserving ($\Delta m_s=0$), the non-radiative decay paths are spin dependent~\cite{ivady2020}. These spin selective processes lead to (i) polarization of the \VB centre in the $m_s = 0$ spin sublevel by optical pumping and (ii) spin-dependent PL emission enabling optical detection of magnetic resonances~\cite{Gottscholl2020}. \\
\indent In the following, we perform ESR spectroscopy of \VB centers hosted in isotopically purified hBN flakes using a scanning confocal microscope operating under green laser excitation with a high numerical aperture objective (NA=0.65) and a photon counting module for PL detection. A microwave excitation is delivered by an external loop antenna and a static magnetic field $B_z$ is applied along the c-axis of hBN, referred to as the $z$ axis, with a permanent magnet placed on a three-axis translation stage. ESR spectra are recorded by monitoring the spin-dependent PL signal while sweeping the frequency of the microwave field. All experiments are performed under ambient conditions.\\
\indent We first analyze the impact of isotopic purification on the hyperfine structure of the \VB center. To this end, we isolate a single magnetic resonance of the spin triplet ground state by applying a weak magnetic field that splits the electron spin sublevels $m_s=\pm1$ via the Zeeman effect [Fig.~\ref{fig1}(a)]. ESR spectra are recorded at low optical and microwave powers to avoid power broadening effects. For \VB centers in h$^{10}$B$^{14}$N, we observe the usual seven-line hyperfine structure of the magnetic resonance, with a characteristic splitting $\mathcal{A}_{zz}(^{14} \rm N)=44.3(2)$~MHz [Fig.~\ref{fig1}(b)]. This structure results from the hyperfine interaction of the \VB electron spin with the three neighboring $^{14}$N nuclei, whose total nuclear spin projection along the $z$ axis takes the values $m_{I}=\{0, \pm 1,\pm2, \pm3\}$, thus leading to seven hyperfine lines~\cite{Gottscholl2020,haykal2021decoherence}. The same experiment carried out for \VB centers in h$^{10}$B$^{15}$N shows a radically different hyperfine structure [Fig.~\ref{fig1}(c)], with only four lines associated to total nuclear spin projections $m_{I}=\{\pm 1/2, \pm 3/2\}$ of the three neighboring $^{15} \rm N$ atoms. In addition the hyperfine splitting is increased to $|\mathcal{A}_{zz}(^{15} \rm N)|=64.1(2)$~MHz, yielding well-resolved ESR lines. This is due to the greater nuclear gyromagnetic ratio of the $^{15}$N isotope ($\gamma_n=-4.3$~MHz/T) compared to $^{14}$N ($\gamma_n=3.07$~MHz/T). Since the strength of the hyperfine coupling scales linearly with $\gamma_n$, the hyperfine splittings are linked by $\left|\frac{\mathcal{A}_{zz}(^{15} \rm N)}{\mathcal{A}_{zz}(^{14} \rm N)}\right|=\left|\frac{\gamma_n(^{15} \rm N)}{\gamma_n(^{14} \rm N)}\right|\sim 1.4$, as observed in our experiments. \\
\indent The boron isotope has also an impact on the magnetic resonance signal. Indeed, while the hyperfine coupling with the six boron atoms located in the second neighboring lattice sites of the \VB center cannot be resolved, this interaction is responsible for the broadening of each ESR line~\cite{haykal2021decoherence}. For the h$^{10}$B$^{15}$N crystal, we obtain a linewidth $\Delta(^{10}\rm B)=44.3(3)$~MHz [Fig.~\ref{fig1}(c)]. By changing the boron isotope to $^{11}$B, the linewidth is increased to $\Delta(^{11}\rm B)=52.9(3)$~MHz [Fig.~\ref{fig1}(d)]. Although the $^{11}$B isotope ($I=3/2$) has a smaller nuclear spin than $^{10}$B ($I=3$), thus inducing fewer hyperfine lines, its nuclear gyromagnetic ratio is three times higher, which results in a greater hyperfine splitting that leads to an increased broadening of the ESR lines~\cite{haykal2021decoherence}. This effect is reproduced by simple simulations of hyperfine spectra that only take into account the effect of the longitudinal components $\mathcal{A}_{zz}$ of the hyperfine interaction (see~SI). These results establish that isotopic purification of hBN with $^{15}$N and $^{10}$B leads to a simplified, well-resolved hyperfine structure with narrower ESR lines, making h$^{10}$B$^{15}$N the optimal host material for future developments of 2D quantum technologies based on \VB centers.\\
\indent We now focus on dynamic polarization of the $^{15}$N nuclei via optical pumping of \VB centers. The polarization mechanism relies on electron-nuclear-spin flip-flop processes induced by the hyperfine interaction, combined with electron spin initialization under optical pumping~\cite{Gao2022,PhysRevLett.102.057403,PhysRevB.87.125207,PhysRevB.102.224101,PhysRevLett.114.247603}. In the following, the polarization of $^{15}$N nuclear spins is studied in a  h$^{10}$B$^{15}$N crystal by recording hyperfine spectra of the high-frequency magnetic resonance of \VB centers, {\it i.e.} $m_s=0\rightarrow m_s=+1$. \\
\indent Typical ESR spectra measured at a magnetic field $B_z=92$~mT are shown in Fig.~\ref{fig2}(a). The relative amplitudes of the hyperfine lines corresponding to the total nuclear spin projections $m_I=+1/2$ and $m_I=+3/2$ become stronger when the optical power increases, which unambiguously indicates optically-induced nuclear polarization. Within the limits of our experimental accuracy, the frequencies of the ESR lines are not shifted, suggesting that only the first $^{15}$N neighbors are efficiently polarized. By fitting the hyperfine spectra with a sum of four Gaussian functions, we extract the area $\mathcal{S}_{m_I}$ of each hyperfine line. The average polarization of each neighboring $^{15}$N nuclear spin is then inferred as
\begin{equation}
\mathcal{P}=(\sum_{m_I}m_I\mathcal{S}_{m_I})/(\frac{3}{2}\sum_{m_I}\mathcal{S}_{m_I}) \ .
\label{nuc}
\end{equation}
As shown in Fig.~\ref{fig2}(b), nuclear polarization increases with the optical excitation power $P_L$ before saturating at $\mathcal{P}_{\rm max}\sim20\%$. We note that this behavior is not related to saturation effects in the optical transition of the \VB center, since the PL signal evolves linearly with $P_L$ in the considered power range.\\
 \begin{figure}[t!]
  \centering
  \includegraphics[width = 8.4cm]{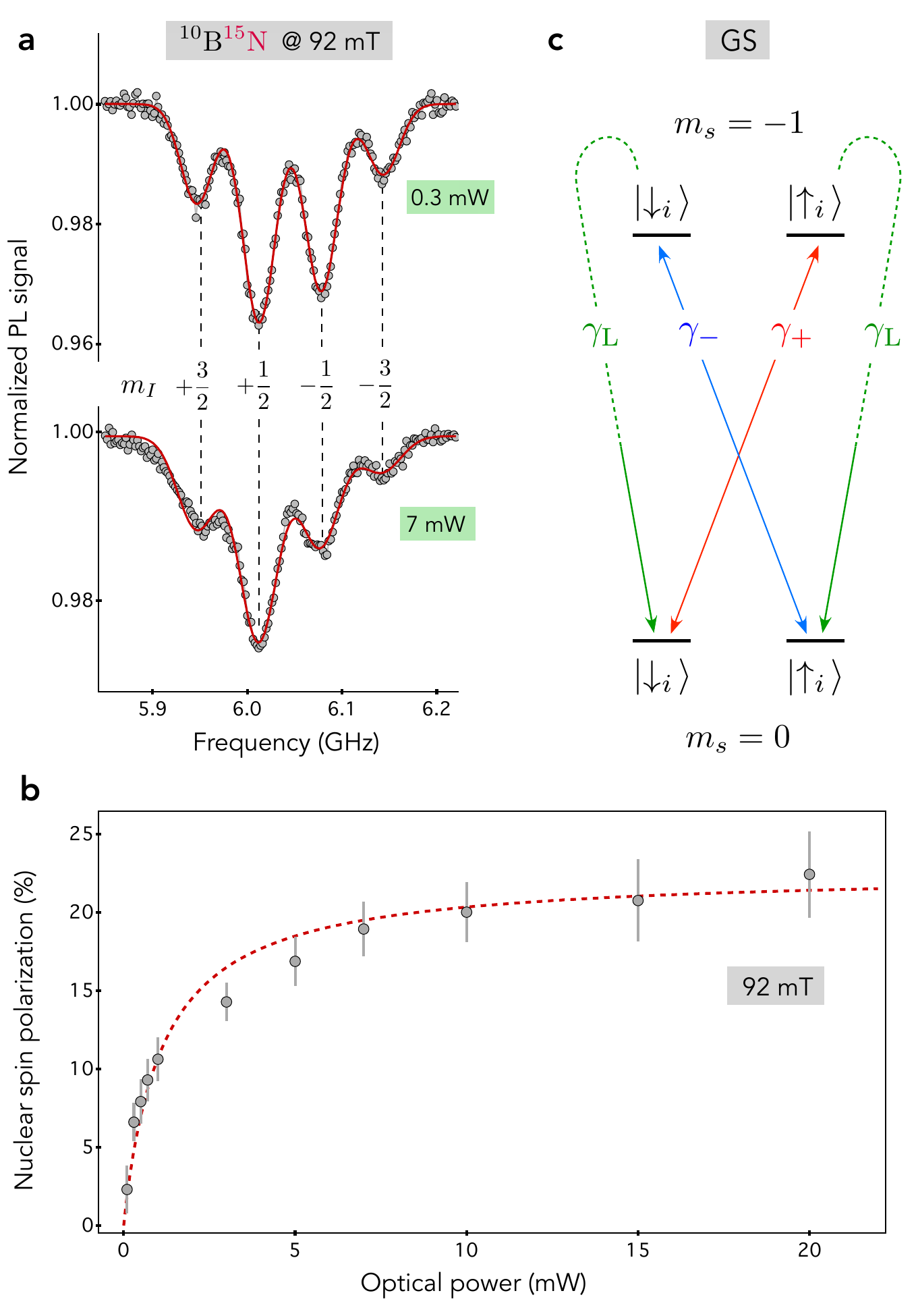}
  \caption{(a) Hyperfine spectra of the high-frequency magnetic resonance recorded at $B_z=92$~mT for $P_L=0.3$~mW (top panel) and  $P_L=7$~mW (bottom panel). Solid lines are data fitting with a sum of four Gaussian functions from which the nuclear spin polarization $\mathcal{P}$ is extracted. (b) $\mathcal{P}$ as a function of $P_L$. The red dashed line is data fitting with a saturation function $\mathcal{P}=\mathcal{P}_{\rm max}/(1+P_L/P_{sat})$, yielding $\mathcal{P}_{\rm max}=22.6(7)\%$ and $P_{sat}=1.1(2)$~mW. (c) Simplified four-level model used to describe the polarization mechanism for a single $^{15}$N nucleus.}
  \label{fig2}
\end{figure}
\indent To understand these results, we introduce the ground state spin Hamiltonian of the \VB center electronic spin coupled by hyperfine interaction with the three nearest $^{15}$N nuclear spins
\begin{equation}
\hat{\mathcal{H}}=D_{\rm g}\hat{S}_z^2 + \gamma_e B_z\hat{S}_z-\sum_{i=1}^{3}\gamma_n B_z\hat{I}_{z}^{(i)}+\underbrace{\sum_{i=1}^{3}\mathbf{\hat{S}} \underline{\underline{\mathcal{A}}}^{(i)}\mathbf{\hat{I}^{(i)}} }_{\hat{\mathcal{H}}_{\rm hf}}  \ ,
\label{Hamilto}
\end{equation}
where $ \underline{\underline{\mathcal{A}}}^{(i)}$ denotes the hyperfine interaction tensor and $\gamma_e\approx 28$~GHz/T is the electronic spin gyromagnetic ratio. The Hamiltonian describing the coupled spin system in the excited state has the same form, albeit with different zero-field splitting parameter and hyperfine tensor. However, we will show below that the excited state has little impact on the nuclear polarization reported in this work. We thus only consider the ground state spin Hamiltonian. For a diagonal tensor, the hyperfine coupling term can be expanded as
\begin{eqnarray*}
\hat{\mathcal{H}}_{\rm hf}=\sum_{i=1}^{3}\mathcal{A}_{zz}^{(i)}\hat{I}_{z}^{(i)} &+&\sum_{i=1}^{3}\mathcal{A}_+^{(i)}\left( \hat{S}_{-}\hat{I}_{+}^{(i)}+\hat{S}_{+}\hat{I}_{-}^{(i)}\right) \\
&+& \sum_{i=1}^{3} \mathcal{A}_-^{(i)}\left( \hat{S}_{+}\hat{I}_{+}^{(i)}+\hat{S}_{-}\hat{I}_{-}^{(i)}\right)  \ ,
\end{eqnarray*}
where $\mathcal{A}_{\pm}^{(i)}=(\mathcal{A}_{xx}^{(i)}\pm\mathcal{A}_{yy}^{(i)})/4$ and $\hat{S}_{\pm}$ (resp. $\hat{I}_{\pm}^{(i)}$) are the electron (resp. nuclear) spin ladder operators. When a magnetic field brings the \VB center near the ground state level anticrossing (GSLAC) [see Fig.~\ref{fig1}(a)], the hyperfine interaction mixes the $m_s=0$ ($\left|0_e \right.\rangle$) and $m_s=-1$ ($\left|-1_e \right.\rangle$) electron spin manifolds. Such a mixing plays a central role in optically-induced nuclear polarization.

The polarization mechanism is first qualitatively analyzed using rate equations within the simple four-level model shown in Fig.~\ref{fig2}(c). We only consider a single $^{15}$N nuclear spin and the eigenstates of the $I_z$ operator are denoted as $\left|\uparrow_i \right.\rangle$ and $\left|\downarrow_i\right.\rangle$. The second term of the hyperfine interaction couples the states $\left|0_e,\downarrow_i \right.\rangle$ and $\left|-1_e,\uparrow_i\right.\rangle$, leading to electron-nuclear-spin flip-flops, which are described by a bi-directional transition rate $\gamma_{\rm +}\propto |\mathcal{A}_{+}^{(i)}|$. In addition, the population in state $\left|-1_e,\uparrow_i \right.\rangle$ can be transferred to $\left|0_e,\uparrow_i \right.\rangle$ via optical pumping with a rate $\gamma_L\propto P_L$. These two combined processes promote spin populations in state $\left|0_e,\uparrow_i \right.\rangle$. This mechanism competes with the third term of the hyperfine interaction, which couples the states $\left|0_e,\uparrow_i \right.\rangle$ and $\left|-1_e,\downarrow_i \right.\rangle$ with a rate $\gamma_{\rm -}\propto |\mathcal{A}_{-}^{(i)}|$, and thus promotes populations in state $\left|0_e,\downarrow_i \right.\rangle$ by optical pumping. In this simple framework, the nuclear polarization along the $z$ axis is inferred by calculating the difference of the steady-state populations in $\left|0_e,\uparrow_i \right.\rangle$ and $\left|0_e,\downarrow_i \right.\rangle$. We obtain a characteristic saturation behavior of the polarization with the optical power, as observed in the experiments [Fig.~\ref{fig2}(b)], with a  maximum polarization at saturation given by [see SI]
\begin{equation}
\mathcal{P}_{\rm max}=\frac{\gamma_+-\gamma_-}{\gamma_++\gamma_-}=\frac{1-|\mathcal{A}_{-}^{(i)}/\mathcal{A}_{+}^{(i)}|}{1+|\mathcal{A}_{-}^{(i)}/\mathcal{A}_{+}^{(i)}|} \ .
\end{equation}

Ab initio calculations of the hyperfine tensor for $^{14}$N nuclei yield $\mathcal{A}_{xx}^{(i)}\sim 47$~MHz and $\mathcal{A}_{yy}^{(i)}\sim 90$~MHz~\cite{Gao2022,ivady2020}. Since the ratio between hyperfine components is isotope-independent, we use these numbers to infer $|\mathcal{A}_{-}^{(i)}/\mathcal{A}_{+}^{(i)}|\sim 0.3$, leading to $\mathcal{P}_{\rm max}\sim 50\%$. Although this value cannot be quantitatively compared to experimental results given the simplicity of the model, our analysis indicates that the polarization efficiency of $^{15}$N nuclei is intrinsically limited by the unequal components $\mathcal{A}_{xx}^{(i)}$ and $\mathcal{A}_{yy}^{(i)}$ of the hyperfine tensor, which results from the symmetry of the \VB center. This situation differs from dynamic nuclear spin polarization experiments performed with other optically-active spin defects hosted in 3D materials such as nitrogen-vacancy (NV) centers in diamond~\cite{PhysRevLett.102.057403,PhysRevB.87.125207,PhysRevB.102.224101} or divacancies in SiC~\cite{PhysRevLett.114.247603}. For these defects, an almost perfect polarization ($\sim 100\%$) of nearby nuclear spins can be achieved by optical pumping because the hyperfine contact interaction is always characterized by $\mathcal{A}_{xx}\sim \mathcal{A}_{yy}$~\cite{PhysRevB.92.115206}, such that $\mathcal{A}_{-}\sim 0$.   

 \begin{figure}[t!]
  \centering
  \includegraphics[width = 8.7cm]{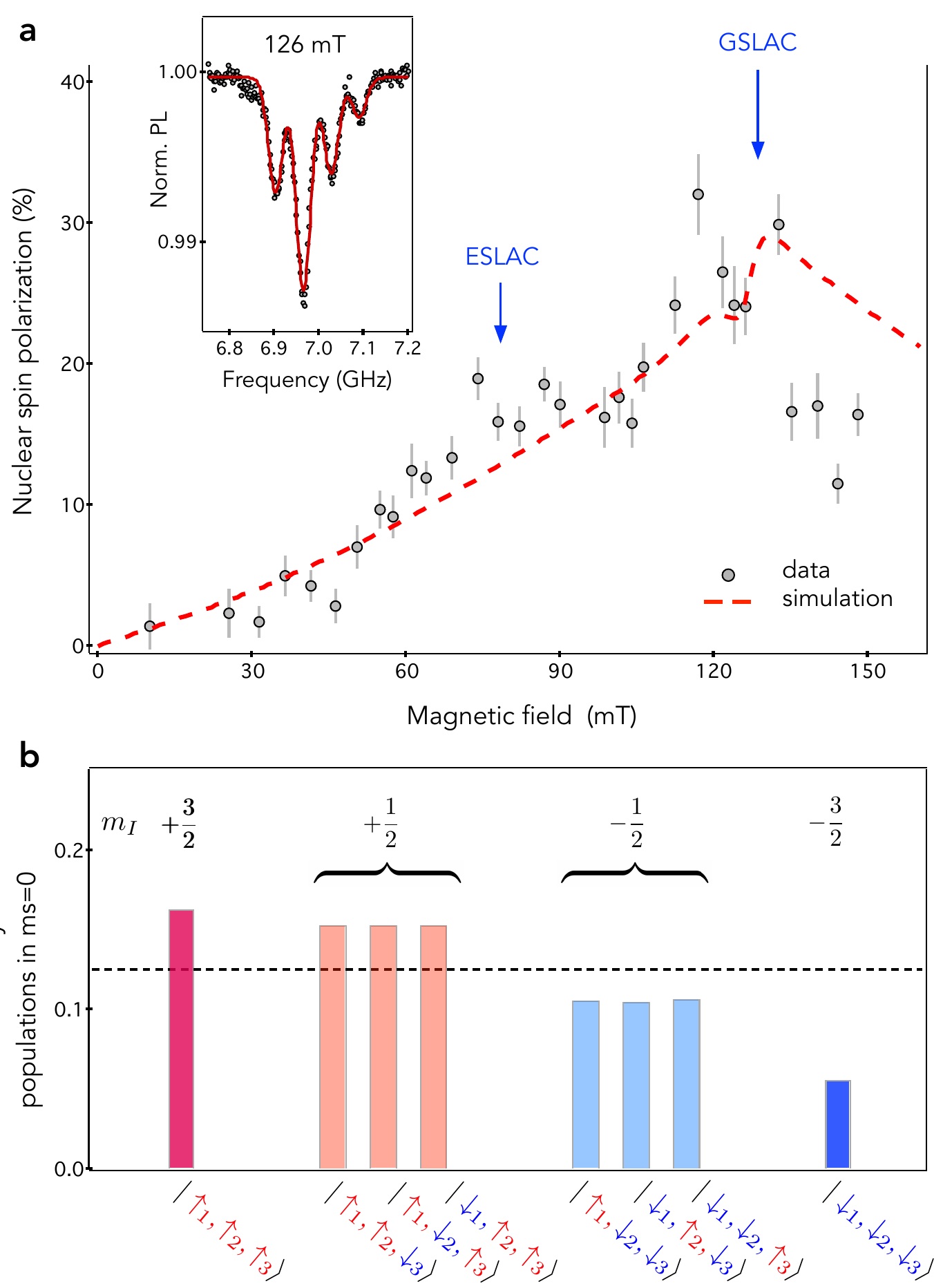}
  \caption{(a) Nuclear spin polarization as a function of the magnetic field $B_z$ for a laser power of $10$~mW. The red dashed line is the result of numerical simulations for a laser pumping rate $\Gamma_L=10$~MHz (see SI). The inset shows the ESR spectrum recorded near the GSLAC. (b) Normalized steady state populations in the $m_s=0$ ground state manifold calculated at $B_z=92$~mT for $\Gamma_L=10$~MHz. The black dashed line corresponds to unpolarized $^{15}$N nuclei.}
  \label{fig3}
\end{figure}

We now study how nuclear polarization evolves with the magnetic field $B_z$. For these experiments, we employ a laser power $P_L=10$~mW and the field alignment is realized (i) by optimizing the PL signal and (ii) by verifying that the average of the two ESR frequencies of the \VB center is equal to the zero-field splitting parameter $D_{\rm g}$. This procedure provides a field alignment accuracy better than $\sim1^{\circ}$ (see SI). As shown in Fig.~\ref{fig3}(a), nuclear polarization increases with the magnetic field before reaching a maximum $\mathcal{P}\sim 30\%$ at the GSLAC ($B_z\sim 124$~mT). The polarization efficiency is however very sensitive to the magnetic field alignment along the $z$ axis [see SI]. Indeed, any field component perpendicular to the \VB quantization axis opens additional depolarization channels by mixing the electron and nuclear spin sublevels via the Zeeman terms of the spin Hamiltonian.

These results are compared to numerical simulations of the nuclear polarization over a wide range of magnetic fields. To this end, we consider the \VB center interacting with the three neighboring $^{15}$N both in the triplet ground state and in the triplet excited state, using the spin Hamiltonian given by Eq.~(\ref{Hamilto}). The nuclear polarization is inferred from the steady-state solution of the master equation, in which decoherence processes, spin conserving optical transitions and spin-dependent transitions to a metastable singlet state are introduced via a Lindblad superoperator~\cite{Gao2022,PhysRevB.102.224101} (see SI). We note that such calculations are drastically simplified for $^{15}$N nuclei compared to $^{14}$N because (i) the total number of nuclear spin states drops from $3^3=27$ to $2^3=8$ in each electron spin manifold and (ii) there is no quadrupolar hyperfine interaction term for a nuclear spin $I=1/2$. Figure~\ref{fig3}(b) shows a typical result of the steady-state populations in the $m_s=0$ ground state manifold calculated for a magnetic field $B_z=92$~mT and a laser pumping rate $\Gamma_L=10$~MHz. The nuclear polarization is then inferred by using Eq.~(\ref{nuc}), where $\mathcal{S}_{m_I}$ is the sum of the populations with a total spin projection $m_I$. A fair agreement is obtained between numerical simulations and experiments [Fig.~\ref{fig3}(a)]. Interestingly, these results indicate that the excited-state level anticrossing (ESLAC) has little impact on nuclear polarization. This is explained by (i) the longer residence time of the \VB center in the ground state when the optical pumping power is well below the saturation of the optical transition, combined with (ii) the low probability of electron-nuclear spin flip-flops within the short excited-state lifetime ($\sim 1$~ns)~\cite{Gottscholl2020}. Achieving ESLAC-assisted nuclear polarization would require the use of much higher optical powers than those used in this work~\cite{ru2023robust}.

To conclude, we have shown that isotope engineering of hBN is a powerful approach to tailor the magnetic resonance signal of \VB spin defects, with optimal properties obtained for the combination of $^{15}$N and $^{10}$B isotopes.  Beyond \VB centers, these methods could also be used to study the isotope-dependent hyperfine structure of other spin defects in hBN, which might help to identify their microscopic structure~\cite{Aharanovich_NatMat2021,Chejanovsky2021,Auburger2021,Stern2022,stern2023quantum}. The well-resolved hyperfine structure of \VB centers hosted in h$^{10}$B$^{15}$N enables unambiguous measurements of optically-induced nuclear spin polarization of neighboring $^{15}$N atoms. These nuclei could be used in future as ancillary quantum memories to enhance the sensitivity of quantum sensors based on \VB centers~\cite{Zaiser2016,Rosskopf2017,Pfender2017} and offer additional ressources for exploring many-body physics in a 2D material platform.\\

\noindent {\it Acknowledgements} - The authors acknowledge Viktor Iv\'ady for fruitful discussions. This work was supported by the French Agence Nationale de la Recherche under the program ESR/EquipEx+ 2DMAG (grant number ANR-21-ESRE-0025), a grant from NanoX in the framework of the ''Programme des Investissements d’Avenir'' (ANR-17-EURE-0009, Q2D-SENS), an Office of Naval Research award N00014-22-1-2582, the program QuantEdu France, and the Institute for Quantum Technologies in Occitanie through the project BONIQs and 2D-QSens.\\

\noindent {\it Note} - During the completion of this work, we became aware of a complementary work studying \VB properties in isotopically-engineered hBN crystals (Chong Zu, private communication), which will appear in the same arXiv posting \cite{gong2023isotope}.\\

\end{document}